\documentclass[pra,aps,10pt,twocolumn,superscriptaddress,longbibliography]{revtex4-1}
\usepackage[utf8]{inputenc}
\usepackage{amsmath}
\usepackage{amssymb}
\usepackage{amsthm}
\usepackage{amsfonts}
\usepackage{enumerate}
\usepackage{bm}
\usepackage{graphicx}
\usepackage{subfigure}
\usepackage{color}
\usepackage{nicefrac}
\usepackage[dvipsnames]{xcolor}
\usepackage[colorlinks]{hyperref}
\usepackage{ulem}
\usepackage[capitalise]{cleveref}

\UseRawInputEncoding

\newcommand{\btheta}{{\boldsymbol{\theta}}}
\newcommand{\bgamma}{{\boldsymbol{\gamma}}}
\newcommand{\bbeta}{{\boldsymbol{\beta}}}
\newcommand{\bX}{{\mathbf{X}}}
\newcommand{\tr}{\mathrm{tr}}
\newcommand{\dg}{^\dagger}
\newcommand{\ket}[1]{\vert#1\rangle}
\newcommand{\bra}[1]{\langle #1\vert}

\newcommand{\nn}{\nonumber}
\newcommand{\h}[1]{\hat{#1}}

\newcommand{\eg}{\textit{e.g.,}~}
\newcommand{\ie}{\textit{i.e.,}~}
\newcommand{\etal}{\textit{et al.}~}

\begin{document}

\title{Surrogate-based optimization for variational quantum algorithms}
\author{Ryan Shaffer}
\thanks{Current affiliation: AWS Quantum Technologies, Seattle, WA 98170, USA. Work done prior to joining Amazon.}
\affiliation{Department of Physics, University of California, Berkeley, CA 94720, USA}
\affiliation{Quantum Algorithms and Applications Collaboratory, Sandia National Laboratories, Livermore, CA 94551, USA}
\author{Lucas Kocia}
\affiliation{Quantum Algorithms and Applications Collaboratory, Sandia National Laboratories, Livermore, CA 94551, USA}
\author{Mohan Sarovar}
\email{mnsarov@sandia.gov}
\affiliation{Quantum Algorithms and Applications Collaboratory, Sandia National Laboratories, Livermore, CA 94551, USA}

\begin{abstract}
Variational quantum algorithms are a class of techniques intended to be used on near-term quantum computers. The goal of these algorithms is to perform large quantum computations by breaking the problem down into a large number of shallow quantum circuits, complemented by classical optimization and feedback between each circuit execution. One path for improving the performance of these algorithms is to enhance the classical optimization technique. Given the relative ease and abundance of classical computing resources, there is ample opportunity to do so. In this work, we introduce the idea of learning surrogate models for variational circuits using few experimental measurements, and then performing parameter optimization using these models as opposed to the original data. We demonstrate this idea using a surrogate model based on kernel approximations, through which we reconstruct local patches of variational cost functions using batches of noisy quantum circuit results. Through application to the quantum approximate optimization algorithm and preparation of ground states for molecules, we demonstrate the superiority of surrogate-based optimization over commonly-used optimization techniques for variational algorithms.
\end{abstract}

\date{\today}
\maketitle

\section{Introduction}
As the quality and scale of quantum information processors (QIPs) increase, the question of whether they can derive some advantage over conventional (classical) computers, even before reaching the fault-tolerant regime, is becoming increasingly important to the field. Hybrid algorithms that utilize quantum and classical computing are perhaps the most promising route to such an advantage, and variational algorithms (VQAs) where the QIP evaluates a parametrized cost function that is then optimized by a classical computer are the prime example of such hybrid algorithms \cite{Magann_review_2021, Cerezo_review_2021}.

Conventional implementations of variational algorithms evaluate a parametrized cost function, $V(\btheta)$, usually representing a parametrized quantum circuit, and then optimize over $\btheta$ using off-the-shelf multi-parameter optimization routines like COBYLA, SPSA, and Nelder-Mead \cite{Lavrijsen_2020, Bonet-Monroig_2021}. Such an approach only minimally exploits the structure of the underlying problem, and moreover, only minimally utilizes the computational power of the classical computing layer. While this approach has been used to demonstrate variational algorithms with a handful of parameters, $|\btheta|\equiv D \leq 10$, it is unclear how its effectiveness and the experimental resources it requires will scale to larger problems, where the number of variational parameters becomes hundreds or thousands.

Motivated by this, we introduce an approach to optimization in variational algorithms that utilizes modern statistical inference tools to reduce the experimental burden when running variational algorithms. This, in effect, moves more of the burden from the QIP to the classical computing layer.
The core of our approach is the construction of a \emph{surrogate model} for the variational cost function from QIP experimental data, and performing optimization with this surrogate model instead of the original data. This is an established approach in optimization theory, and surrogate-based optimization (SBO) has found uses in applications where the optimization cost function is difficult to evaluate due to paucity of data or computational expense \cite{Queipo_2005}.
There are a variety of techniques for learning a surrogate model from data, including spline-based fitting, kriging, and neural network models \cite{Forrester_2008}. In this work we demonstrate SBO for VQAs using local kernel approximation techniques. Kernel approximation is particularly useful for building surrogate models for variational quantum circuits for several reasons: (i) the resulting models are explicitly smooth and smooth out unavoidable shot noise in quantum circuit measurements,  (ii) the models can be learned with \emph{batches} of circuit outputs, which has practical advantages for quantum computing platforms where circuit loading incurs latency, and (iii) the models allow numerically efficient computation of $V(\btheta)$ and its derivatives, thus enabling optimization by scalable gradient-based algorithms.
Intuitively, surrogate models based on a kernel approximation can be seen as explicitly taking advantage of the fact that the underlying variational cost function is smooth ($V(\btheta)\in C^{\infty}$), and thus its value at $\btheta$ indicates its value in its neighborhood. We couple this local surrogate model with an adaptive optimization procedure to efficiently find local optima of the variational cost function.

There have been several recent efforts to develop custom optimizers for VQAs, including:
variations of stochastic gradient descent that adapt the number of experimental circuit evaluations (shots) to manage the tradeoff between cost function and gradient estimation quality and experimental burden \cite{Kubler_2020, Sweke_2020, Gu_2021},
techniques based on Bayesian optimization \cite{self2021variational, Tamiya_SGLBO_2021, Iannelli_NoisyBayesianOptimization_2021},
and machine learning-based optimization approaches for specific VQAs \cite{Khairy_2020}.
Most relevant to this work is the study of Sung \etal \cite{Sung_2020}, which in the framework of SBO, developed local quadratic models based on experimental data and coupled this with a trust-region optimization algorithm. Our work expands on this result by considering more general, non-parametric surrogate models that are designed to be valid over larger regions in parameter space, where the quadratic model might break down.
We note that a related approach based on Gaussian process surrogate models has recently been proposed by Mueller \etal \cite{Mueller_AcceleratingNoisyVQEOpt_2022}.

In the following, we introduce our optimization algorithm (\cref{sec:approach}), analyze its theoretical properties and hyperparameter choices (\cref{sec:theory}), and present several numerical illustrations of the approach, including comparisons to conventional variational optimization algorithms (\cref{sec:eg}). Finally, we conclude with a summary and discussion of possible extensions of our approach (\cref{sec:disc}).

\section{Surrogate-based optimization}
\label{sec:approach}
The goal of the classical computing layer in quantum variational algorithms is to compute
\begin{align}
	\min_\btheta V(\btheta)
\end{align}
and, often, also the argument that attains this minimum. Here, $\btheta = (\theta_1, ... \theta_D) \in [0,2\pi)^D$ are parameters that dictate the variational quantum circuit ansatz for the problem, $V(\btheta): [0,2\pi)^D \to \mathbb{R}$ is the variational cost function, which is related to the parametrized circuit, $\h{U}(\btheta)$, acting on $n$ qubits: $V(\btheta) = \tr [\h{O} \h{U}(\btheta) \h{\rho}_0 \h{U} \dg(\btheta)]$, for some initial $n$-qubit state $\h{\rho}_0$ and observable $\h{O}$. This quantum expectation must be estimated using many measurements on the circuit output. To do so, we first decompose the observable into a sum of non-commuting operators, $\h{O} = \sum_{i=1}^\nu \alpha_i \h{o}_i$, with $[\h{o}_i, \h{o}_j]\neq 0$ for $i\neq j$. For all practical VQAs, $\nu = \mathcal{O}(\textrm{poly}(n))$. Then, writing the circuit output as $\h{\rho}(\btheta)\equiv \h{U}(\btheta)\h{\rho}_0 \h{U}\dg(\btheta)$,
\begin{align}
V(\btheta) &= \sum_{i=1}^\nu 	\alpha_i \tr(\h{o}_i \h{\rho}(\btheta)) = \sum_{i=1}^\nu \alpha_i \mathbb{E}\{ \bX^i(\btheta) \},
\end{align}
where $\bX^i(\btheta)$ is a random variable distributed as $p^i(\btheta)$ that represents the outcome of measuring $\h{\rho}(\btheta)$ in the eigenbasis of $\h{o}_i$. In practice, the expectation in the final expression is estimated using a sample mean of a number of \emph{shots} (executions of the circuit at $\btheta$ and measurements in one of the $\nu$ bases). That is, one takes $K_i$ measurements of $\bX^i(\btheta): X^i_1(\btheta), ..., X^i_{K_i}(\btheta)$, and approximates $\mathbb{E}\{\bX^i(\btheta)\}\approx \frac{1}{K_i}\sum_{j=1}^{K_i} X^i_j(\btheta)$. The total number of shots, or circuit executions, necessary to form an estimate of the cost function at a given parameter value,
\begin{align}
\tilde{V}(\btheta) = \sum_{i=1}^\nu \alpha_i \left( \frac{1}{K_i}\sum_{j=1}^{K_i} X^i_j(\btheta)\right)
\end{align}
is $\mathcal{K} = \sum_{i=1}^\nu K_i$.

Since $V(\btheta)$ must be estimated from a finite number of measurement results, the resulting optimization landscape is noisy and becomes increasingly so as the number of available measurements, $\mathcal{K}$, decreases. This is the impact of so-called quantum \emph{shot noise} (the irreducible uncertainty of quantum systems that results in indeterminate measurement outcomes in general) on the variational optimization problem. The poor performance of most optimization algorithms in such noisy landscapes places a burden on the QIP to produce as many measurements as possible to increase the accuracy of this expectation estimate, and therefore the smoothness of $\tilde{V}(\btheta)$. In addition to this shot noise, in present and near-future generations of noisy intermediate scale quantum (NISQ) devices \cite{Preskill_2018} there are other sources of noise coming from poor control, measurement, and isolation (decoherence) that produce distortions of the underlying probability distribution over measurement outcomes; \ie $p^i(\btheta) \to \tilde{p}^{\,i}(\btheta)$. We do not directly address this source of noise, although we note that several error mitigation techniques have been developed to address this problem, \eg \cite{Temme_Bravyi_Gambetta_2017, Endo_Benjamin_Li_2018,Kandala_2019,Czarnik_2021}, and they can be used in tandem with our optimization approach to achieve some degree of robustness to both sources of noise (shot noise and decoherence).

\begin{figure*}
\centering
  \includegraphics[width=0.9\textwidth]{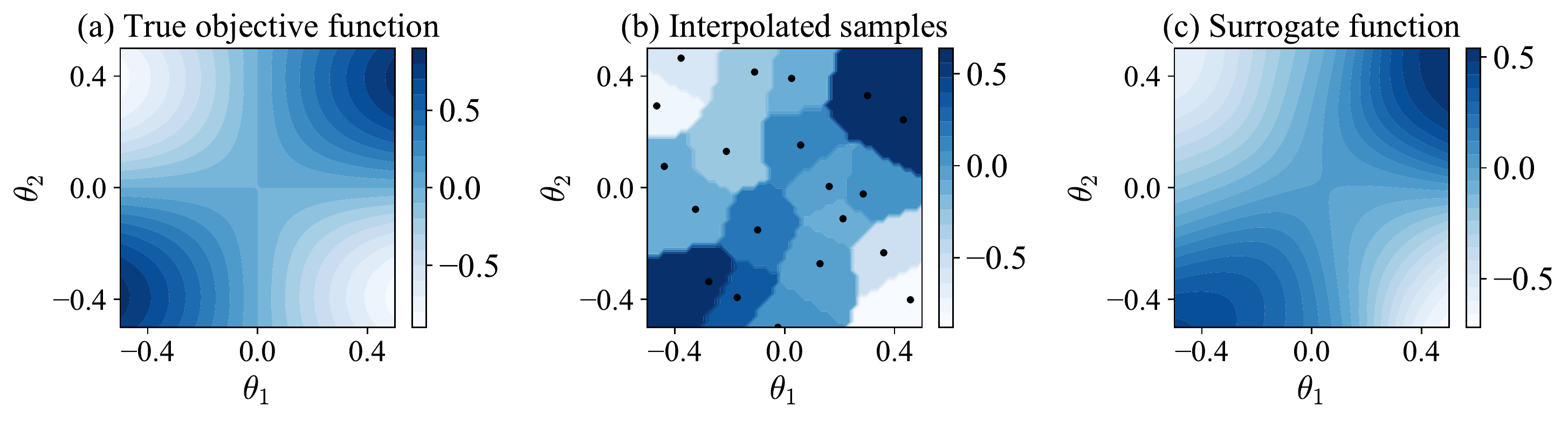}
\caption{An illustration of a local patch of (a) a true objective function $V(\btheta)$ with dimension $D=2$ where $\btheta = (\theta_1, \theta_2)$, (b) interpolated samples $\tilde{V}(\btheta)$ using $\mathcal{K}=100$ shots at each of the $\tau=20$ sample points, and (c) surrogate function $W(\btheta)$ constructed using a Gaussian kernel $\kappa(\btheta, \btheta_j) = \exp(\nicefrac{-||\btheta - \btheta_j||^2}{2\sigma})$.}
\label{fig:surrogate}
\end{figure*}

We now introduce the concept of a local surrogate model to $V(\btheta)$. This is a function $W: \Theta \to \mathbb{R}$ that is an approximation of $V(\btheta)$ in a local \emph{patch}, $\Theta \subset [0,2\pi)^D$. We demand that this surrogate model must be (i) smooth and (ii) efficient to evaluate on a classical computer, requiring no additional measurements from a QIP than those required to construct it. In this work, we construct such a surrogate model using a kernel approximation; \ie
\begin{align}
	W_\Theta(\btheta) = \sum_{j=1}^\tau \tilde{V}(\btheta_j) \kappa(\btheta,\btheta_j),
	\label{eq:kernel_W}
\end{align}
where $\tilde{V}(\btheta_j)$ are standard estimates of $V(\btheta)$ (constructed using $\mathcal{K}$ shots) at $\tau$ distinct \emph{sample points} $\btheta_j \in \Theta$, and $\kappa(\cdot,\cdot)$ is a \emph{kernel function}. Note that the subscript on $W_{\Theta}$ serves to remind us that the surrogate model is valid in some local patch of parameter space, since it is formulated based on data from that local patch.

The choice of $\kappa$ determines most of the properties of kernel-based surrogate models. In this work, we choose a Gaussian kernel, $\kappa(\btheta, \btheta_j) = \exp(\nicefrac{-||\btheta - \btheta_j||^2}{2\sigma})$, for two reasons. First, it is a simple kernel with only one free parameter, $\sigma$, that can be set in a data-driven manner, as we show below. And second, its form allows for easy analytic evaluation of derivatives of $W_\Theta(\btheta)$, which is a useful property for gradient-based optimization of $W(\btheta)$. 

It is known that this kernel can result in a systematic bias \cite{Wendland_2004}. 
In the context of VQAs, this is often manifest in an ``offset'' of the kernel-produced variational cost function values from the experimental values. 
However, given the prevalence of systematic noise in experimental measurements on current quantum hardware, there is frequently no gain to be made from expending computational resources to get the true experimental surface because it is offset already, and only relative magnitudes matter for optimization. 
Moreover, in applications where the goal is finding the parameter argument of the minimal objective function, the offset is irrelevant. 
Finally, in applications where the minimal variational cost function value is desired, it is often possible to fix the offset, both from experimental noise and the kernel, by appealing to special cases when the parameter values simplify the objective function to known values, and shifting the offset of the full surface accordingly.

\cref{fig:surrogate} provides an illustration of a true objective function $V(\btheta)$, interpolated samples $\tilde{V}(\btheta_j)$, and surrogate function $W_\Theta(\btheta)$ constructed using a Gaussian kernel.

\subsection{Adaptive optimization}
As described above, the kernel-based surrogate model is learned over a local patch $\Theta$. In order to find a local optimum of $V(\btheta)$, we couple this construction with an adaptive optimization procedure that we describe in this section.

We begin with an initial seed for the variational parameters, $\btheta^{(0)}$, and define a local patch around it as a $D$-dimensional hypercube of length $\ell$: $\Theta^{(0)} = \cup_{m=1}^D [\theta^{(0)}_m - \nicefrac{\ell}{2}, \theta^{(0)}_m+\nicefrac{\ell}{2}]$. Then we randomly sample $\tau$ points in this patch, execute variational circuits defined by each of those sample points, and use the resulting data to form estimates $\tilde{V}(\btheta_1), ... , \tilde{V}(\btheta_\tau)$. We assume for simplicity that each of the estimates $\tilde{V}(\btheta_j)$ is formed using $\mathcal{K}$ shots, \ie $\mathcal{K}$ does not depend on $j$, although this is not an essential assumption. The $\tau$ samples of $\btheta_j$ are sampled sparsely in $\Theta^{(0)}$; to achieve this in practice, we use Latin hypercube sampling over $\Theta^{(0)}$ to choose each $\btheta_j$. The number of samples $\tau$ and the patch ``size'' $\ell$ are important parameters; we develop heuristics for choosing their values and study their scaling with $n$ and $D$ in \cref{sec:theory} (and also in \cref{app:bound}).
The estimates $\tilde{V}(\btheta_j)$ are then used to formulate a surrogate model $W_{\Theta^{(0)}}$ for $V(\btheta)$ on the patch $\Theta^{(0)}$, as defined in \cref{eq:kernel_W}.

Given $W_{\Theta^{(0)}}(\btheta)$, we perform optimization over this (explicitly smooth) function over the local domain $\Theta^{(0)}$. We do not specify the method to use for this optimization. However, given a smooth objective and easily computable gradients, gradient-based optimizers that incorporate parameter constraints (since the optimization should only be over $\Theta^{(0)}$) are well-suited for this task. In practice, it may be helpful to optimize over a slightly smaller domain to avoid edge effects in the kernel approximation; \ie
\begin{align}
	\min_{\btheta \in \Theta^{(0)}_\epsilon} W_{\Theta^{(0)}}(\btheta),
\end{align}
with $\Theta^{(0)}_\epsilon = \cup_{m=1}^D [\theta^{(0)}_m - \nicefrac{(\ell-\epsilon)}{2}, \theta^{(0)}_m+\nicefrac{(\ell-\epsilon)}{2}]$. The argument that achieves the above minimum defines the center of the next patch, $\btheta^{(1)}$, and this process is repeated.

We refer to the process above as one
\textit{iteration} of the optimization run. Each iteration thus requires $\mathcal{K} \tau$ shots. We perform a fixed number of iterations $M$, giving a total of $\mathcal{K} \tau M$ shots in a full optimization run.
To assist in the convergence of the optimization run, we linearly increase $\epsilon$ from some initial (small) value
$\epsilon_i$
in the first iteration to a value near $\ell$ in the final iteration.

If the minimum $\btheta^{(i+1)}$ found after iteration $i$ falls within the interior of the current patch $\Theta^{(i)}$, \ie if $\btheta^{(i+1)} \in \Theta^{(i)}_{\epsilon_\textrm{int}}$ for some small $\epsilon_\textrm{int} \sim \nicefrac{\ell}{20}$ which excludes the boundary of the patch, then we add the minimum $\btheta^{(i+1)}$ to a list of local minima $\Theta_{\textrm{minima}}$. After completing $M$ iterations, we calculate the final estimated optimum $\btheta_{\textrm{opt}}$ by taking the coordinate-wise mean of all of the elements of $\Theta_{\textrm{minima}}$ that fall within a distance $\ell - \epsilon_f$ (for $\epsilon_f \sim \nicefrac{\ell}{2}$) of the minimum $\btheta^{(M)}$ found in the final iteration; \ie for $\Theta^{(M)}_{\epsilon_f,\textrm{minima}} = \Theta_{\textrm{minima}} \cap \Theta^{(M)}_{\epsilon_f}$,
\begin{align}
    \btheta_{\textrm{opt}} &=
    \frac{1}{\left| \Theta^{(M)}_{\epsilon_f,\textrm{minima}} \right|}
	\sum_{
	   \btheta \in \Theta^{(M)}_{\epsilon_f,\textrm{minima}}
	} \btheta.
\end{align}

\begin{figure}[t]
\centering
  \includegraphics[width=.49\columnwidth]{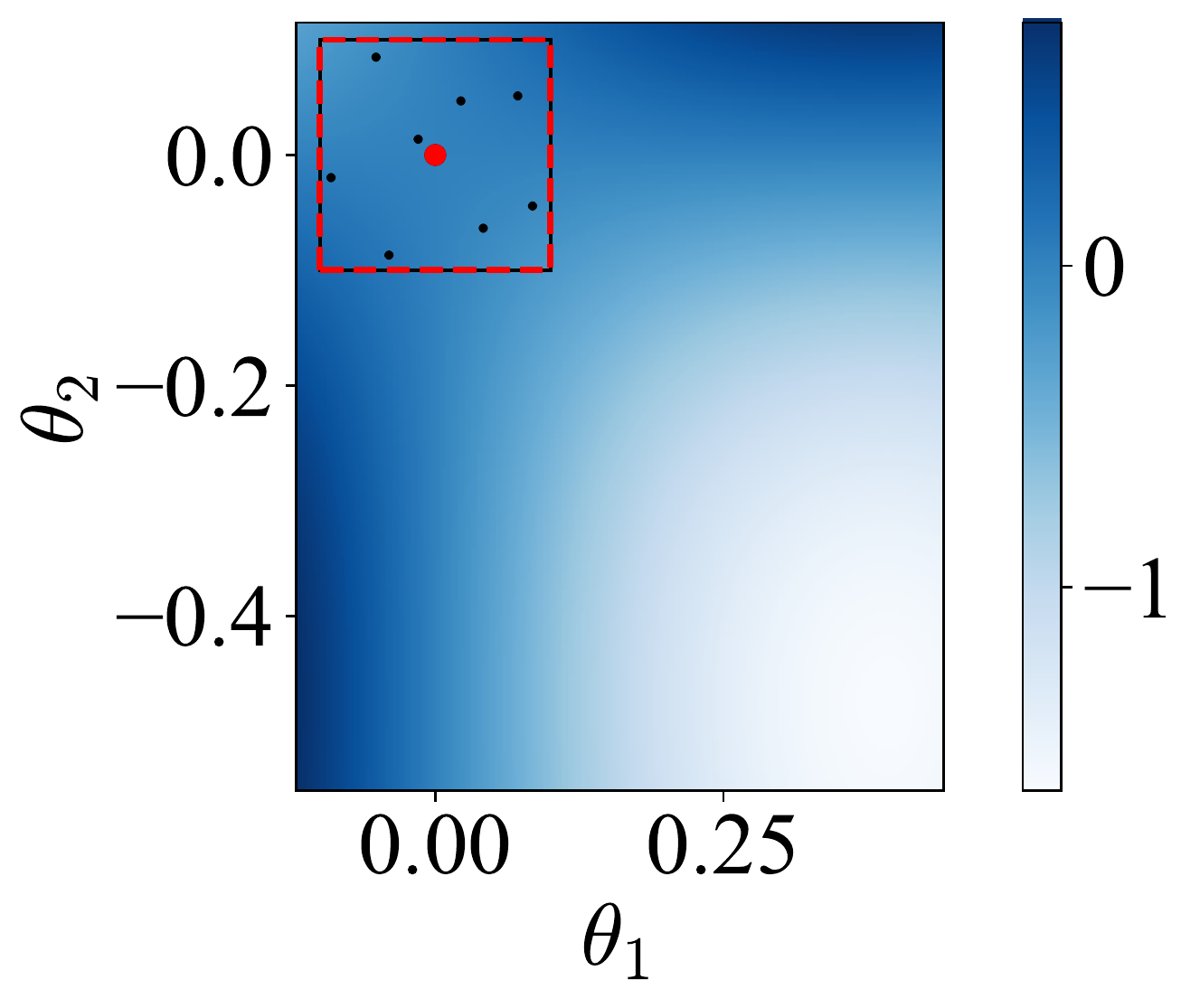}
  \includegraphics[width=.49\columnwidth]{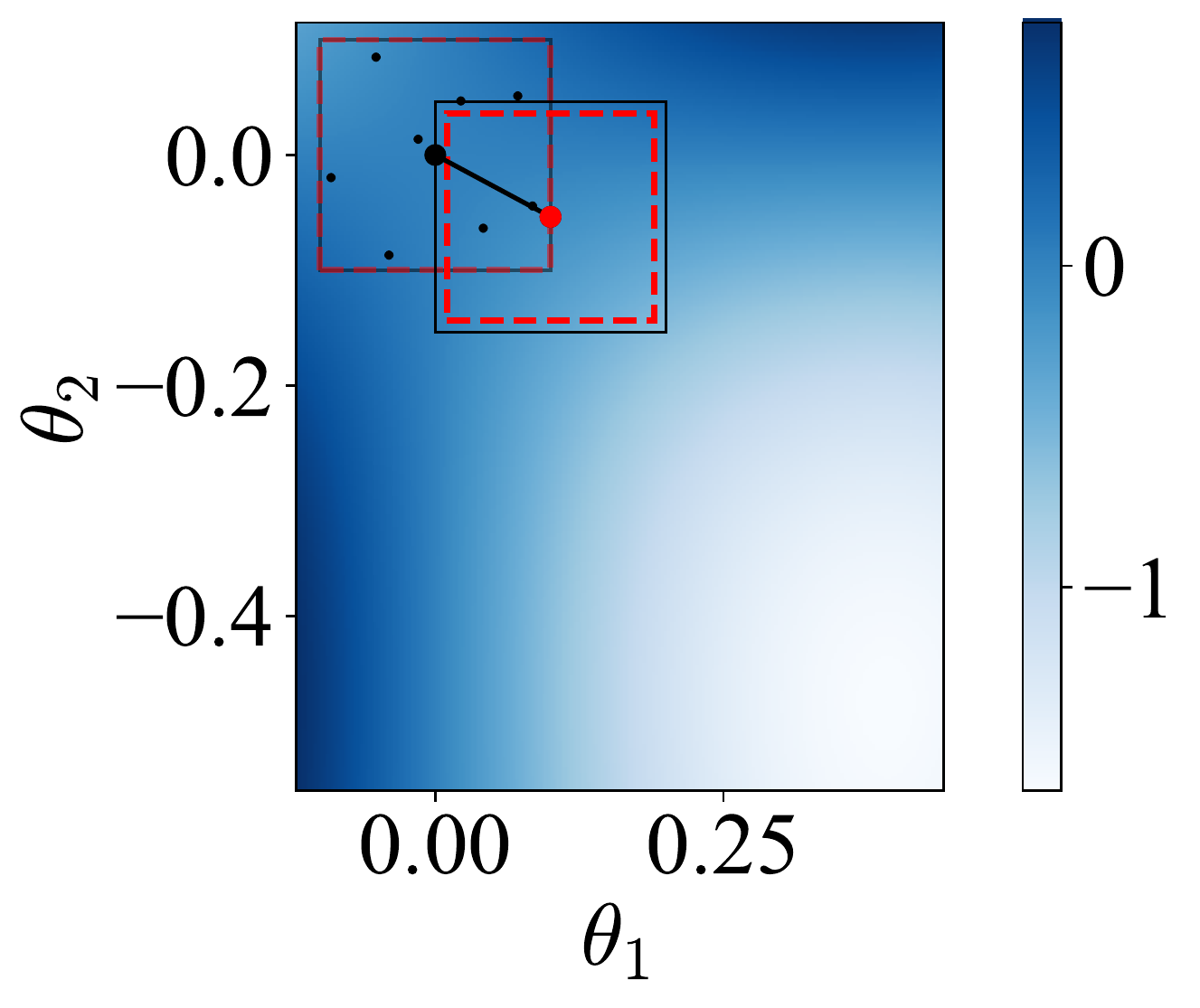}
  \includegraphics[width=.49\columnwidth]{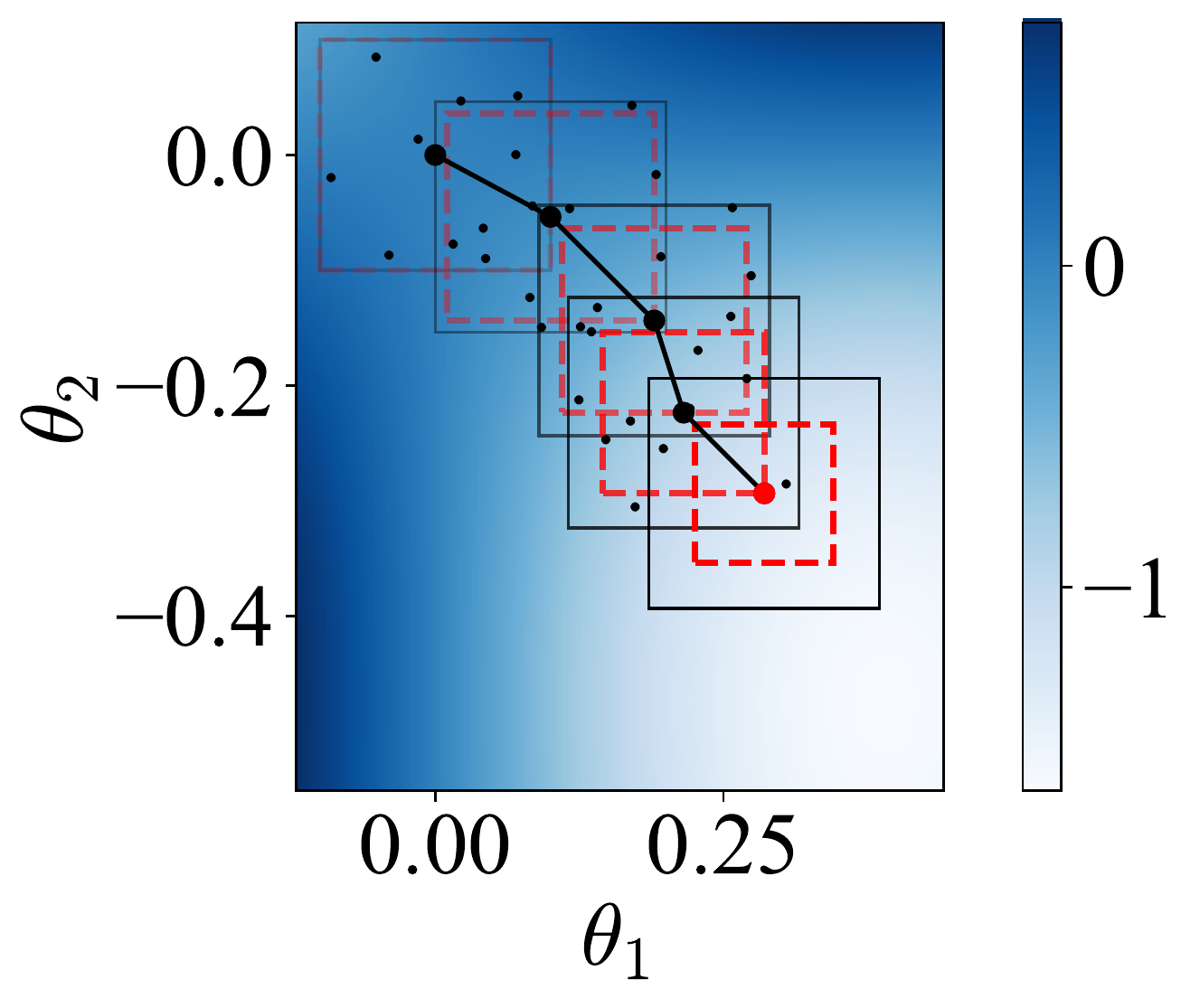}
  \includegraphics[width=.49\columnwidth]{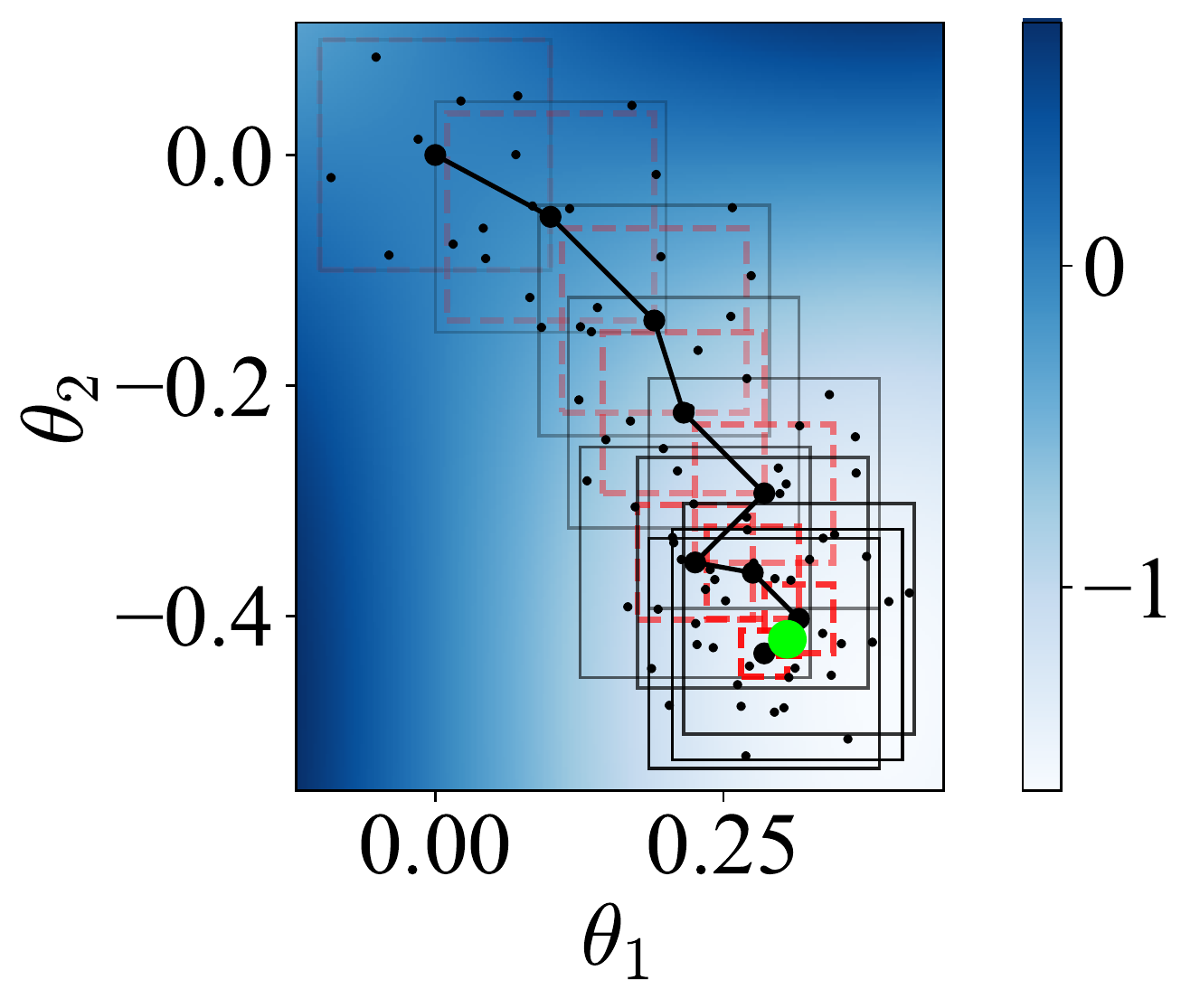}
\caption{A graphical description of the adaptive surrogate-based optimization procedure on a two-dimensional objective function surface parameterized by $\btheta = (\theta_1, \theta_2)$, showing snapshots of an optimization run during the first iteration (upper left), after the first iteration (upper right), after the fourth iteration (lower left), and at the completion of the run after $M=10$ iterations (lower right).
The larger, connected points mark the patch centers $\btheta^{(i)}$ of each iteration, with the red point indicating the center of the most recent patch.
The smaller, unconnected points mark the locations of the $\tau = 8$ samples taken during each iteration.
The solid black rectangles mark the boundaries of the sampling region $\Theta^{(i)}$ for each iteration, where each side has fixed length $\ell = 0.2$.
The dashed red rectangles mark the boundaries of the optimization region $\Theta_\epsilon^{(i)}$ for each iteration, where $\epsilon$ is linearly increased from 0 to $\ell = 0.2$ over the course of the optimization run.
The green point in the final plot (lower right) marks the final estimated optimum $\btheta_{\textrm{opt}}.$}
\label{fig:schematic}
\end{figure}

\cref{fig:schematic} provides a graphical description of the surrogate-based adaptive optimization approach described above.

We note that the optimization approach is decoupled from the surrogate model. Although we have found that the adaptive optimization detailed above is effective, it is by no means unique or optimal. It is possible to modify it or even replace it with another approach while keeping the surrogate model idea intact. In particular, it is likely advantageous to incorporate a memory element that inclues information from previous patches into the decisions made at the current patch -- this is a promising area for future study.

\section{Convergence and hyperparameter choices}
\label{sec:theory}

In this section we discuss practical considerations for choosing various hyperparameters of SBO and the adaptive optimization technique described in \cref{sec:approach}.

\textit{Optimization adjustments $\epsilon_i$, $\epsilon_\textrm{int}$, $\epsilon_f$.} These parameters are used primarily to avoid boundary effects near the edges of each patch region, since during each iteration we sample from only the interior of the patch. In this work, we have used $\epsilon_i = 0$, $\epsilon_\textrm{int} = \nicefrac{\ell}{20}$, and $\epsilon_f = \nicefrac{\ell}{2}$ with good results.

\textit{Measurement shots per measurement basis per sample point, $K$.} The choice of $K$ will be primarily driven by experimental considerations. Larger $K$ is always better since it will reduce shot noise and therefore improve the accuracy of the surrogate model, and in turn the performance of the optimization, but at the cost of increased experimental demands (especially run time).
As we shall demonstrate in the next section, one of the advantages of constructing a surrogate model is an increased robustness of optimization performance to shot noise, and thus SBO can alleviate the experimental burden without sacrificing optimization performance.

\textit{Patch size $\ell$ and sample points per patch $\tau$.} These parameters are intimately related. Intuitively, the larger the patch size, $\ell$, the larger the number of sample points per patch, $\tau$, will need to be in order for the surrogate model to be accurate to the true cost function $V(\btheta)$. Since $\tau$ is closely tied to experimental resources, we find it most useful to think in terms of keeping $\tau$ fixed at a constant, and varying $\ell$. In practice, especially in the near-term, experimental constraints such as device instability and access constraints will dictate how large $\tau$ can be, and therefore we think of it as a fixed parameter, independent of variational problem parameters such as $n$ and $D$. In all of our numerical experiments, including the ones reported in the next section, we have kept $\tau \sim 20$.

Given a fixed, constant $\tau$, the choice of $\ell$ is dictated by the need to accurately capture the shape of the objective function $V(\btheta)$ over the patch in each of the $D$ parameter dimensions. A conservative way to ensure that a fixed number of samples captures the objective function is to demand that this function varies minimally within the patch -- \ie to choose $\ell$ such that there is likely no more than one critical point of $V(\btheta)$ in any $\ell^D$ volume in parameter space. In \cref{app:bound} we study the number of critical points in a general variational cost function and based on a loose bound, derive the scaling $\ell = \Omega(\nicefrac{1}{\text{poly}(D,n)})$. For the empirical studies reported in this paper, we have found that patch sizes in the range $0.1 \le \ell \le 0.2$ worked well for QAOA problems with $n \le 12$ and $p \le 7$ ($D\leq 14$), as well as VQE problems with $n \le 8$ and $D \le 8$, using $K \sim 100$. 

Perhaps the most robust solution to determining $\ell$ is to employ an adaptive method that dynamically adjusts $\ell$ along the optimization path according to a quality of fit metric. This would be possible by moving to a trust-region framework for the optimization \cite{Nocedal_Wright_2006}.

\textit{Surrogate model parameters}. The procedure used to construct the surrogate model will typically have parameters to set. In this work we only consider kernel-based surrogate models, and specifically an isotropic Gaussian kernel $\kappa(\btheta, \btheta_j)$, which has one parameter, the \textit{Gaussian bandwidth $\sigma$.} Intuitively, this parameter describes the volume over which one sample data point influences the behavior of the surrogate model. There is a rich literature on Gaussian kernels and their use in approximation, regression, and smoothing, and as a result, many data-driven heuristics exist for choosing $\sigma$. In practice, we have observed good performance using the Silverman bandwidth heuristic \cite{Silverman_1986} 
$\sigma = \left[ \nicefrac{4}{\tau(D+2)} \right] ^ {\nicefrac{1}{(D+4)}}$,
where $\tau$ is the number of sample points in the current patch.

\section{Illustrations}
\label{sec:eg}
In this section we demonstrate our SBO procedure on some model VQAs through numerical simulation. We also compare optimization performance with one of the most commonly used and recommended optimization methods for VQAs, simultaneous perturbation stochastic approximation (SPSA) \cite{Spall_1998}. SPSA is designed to find optima in the presence of noise in the objective function. Key to its popularity in the resource-constrained setting of VQAs is the fact that it estimates gradients using only two evaluations of the (multiparameter) objective function.

\begin{figure*}[t!]
\centering
  \hspace{-.65\columnwidth}
  \raisebox{-0.95\height}{
    \includegraphics[width=.65\columnwidth]{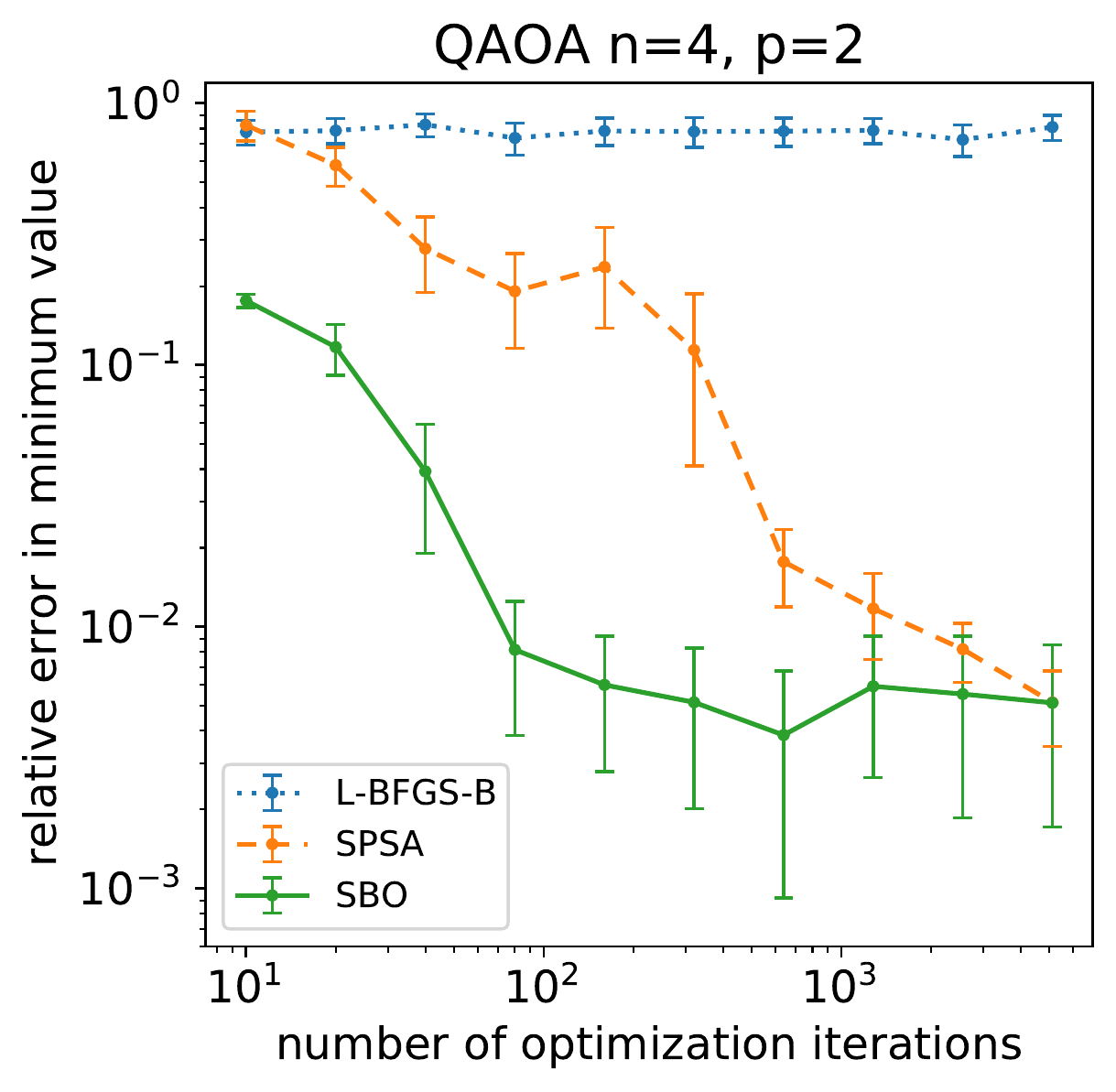}
  }
  \hspace{-.7\columnwidth}\mbox{(a)}\hspace{.65\columnwidth}
  \raisebox{-0.95\height}{
    \includegraphics[width=.65\columnwidth]{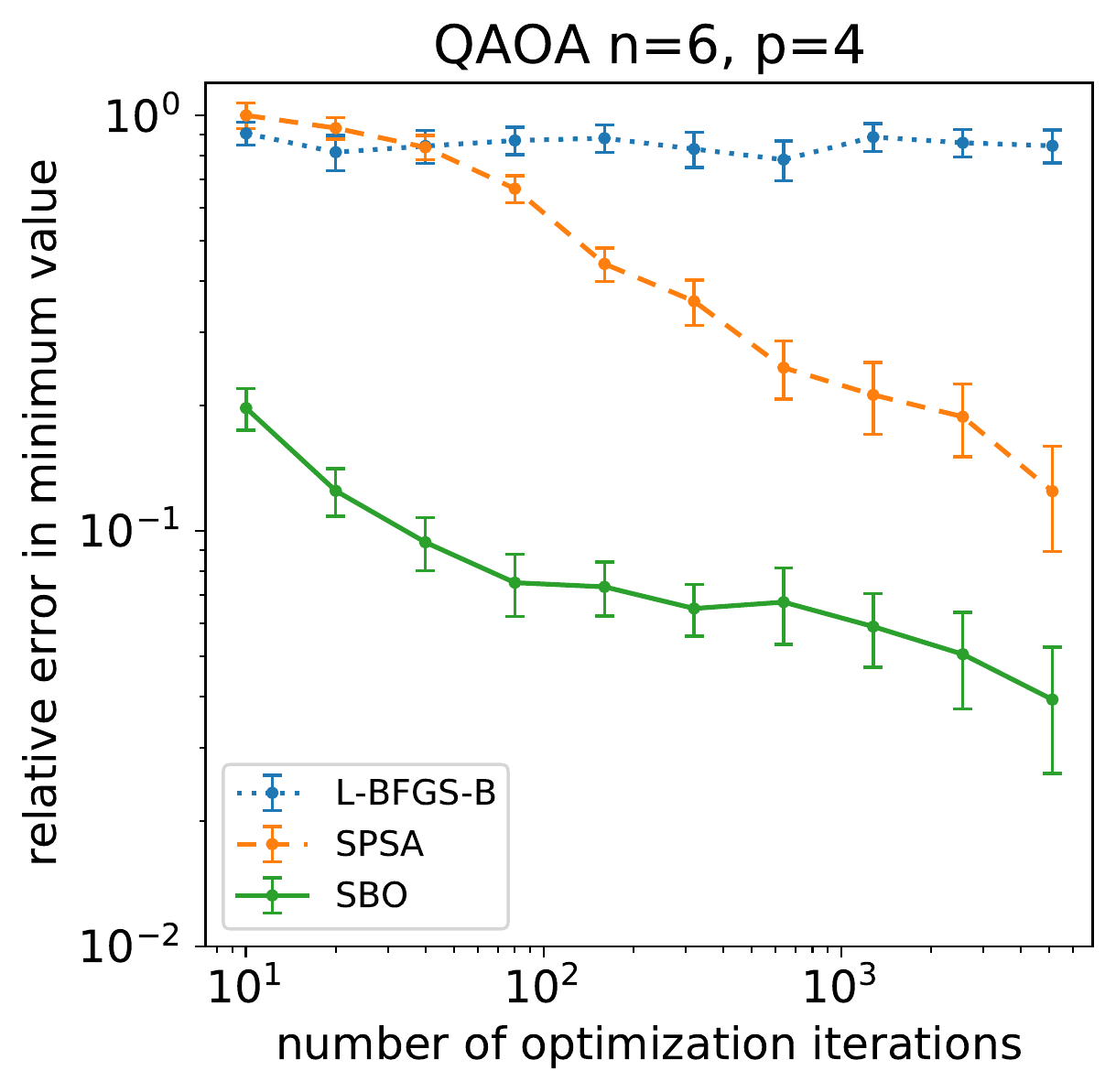}
  }
  \hspace{-.7\columnwidth}\mbox{(b)}\hspace{.65\columnwidth}
  \raisebox{-0.95\height}{
    \includegraphics[width=.65\columnwidth]{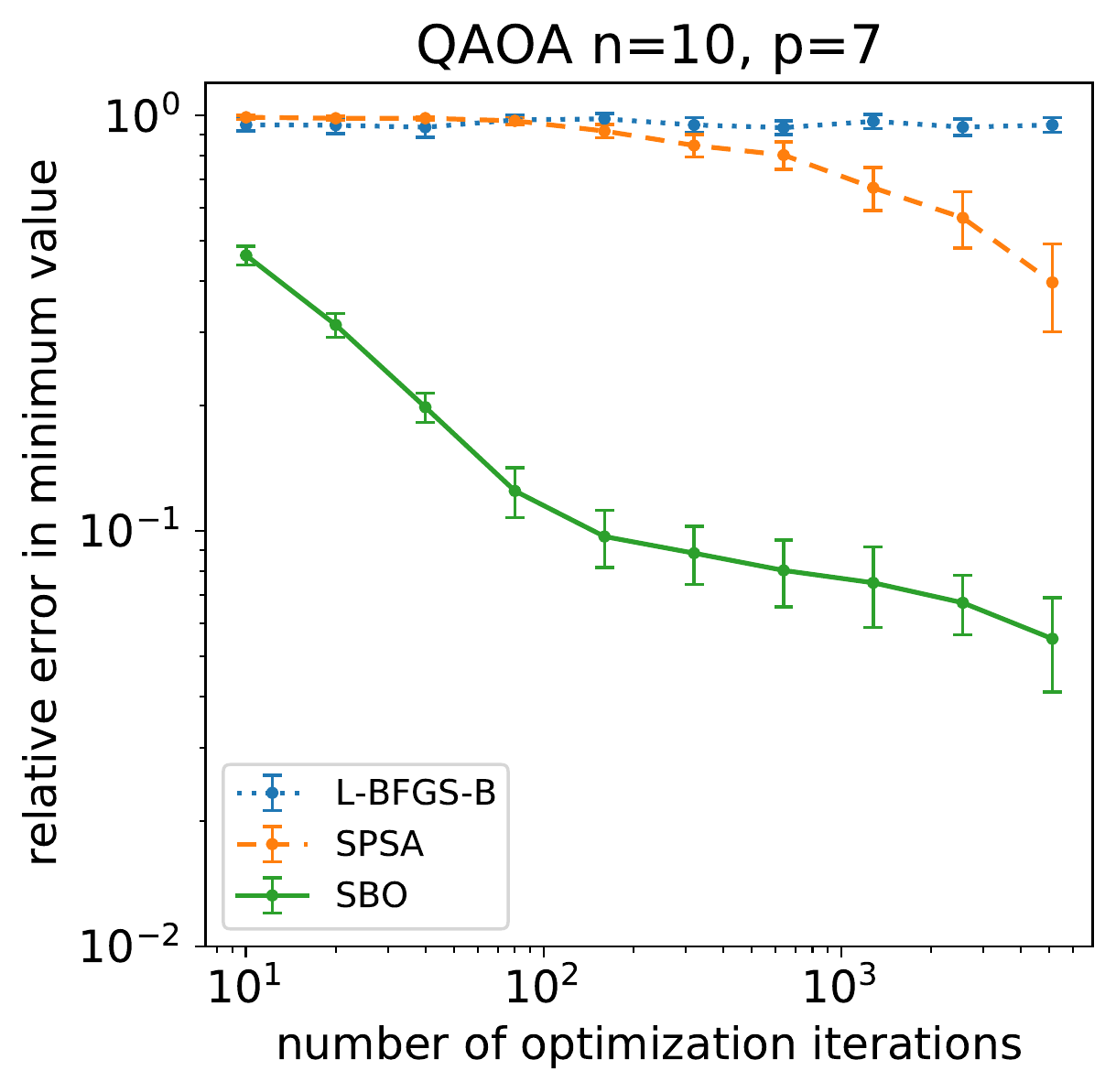}
  }
  \hspace{-.7\columnwidth}\mbox{(c)}
\caption{
    Performance of L-BFGS-B, SPSA, and Gaussian kernel-based SBO optimization runs on QAOA applied to unweighted MaxCut problems of various sizes using an ideal simulator which has only shot noise (and no other errors).
    Each plot displays the results of running $p$-layer QAOA on a single randomly-generated connected graph with $n$ vertices.
    The $x$-axis represents the number of optimization iterations $M$. The $y$-axis represents the relative absolute error achieved by the optimization run, \ie $\left| 1 - V_{QAOA}(\bgamma_{\textrm{opt}}, \bbeta_{\textrm{opt}}) / V_{QAOA,min} \right|$, where $\bgamma_{\textrm{opt}}, \bbeta_{\textrm{opt}}$ are the optimal coordinates obtained by the optimization run and $V_{QAOA,min} = \min_{\bgamma,\bbeta} V_{QAOA}(\bgamma,\bbeta)$ is the true optimum of the objective function.
    Each run uses $K \tau = 5000$ total shots per iteration, where $\tau$ is the number of sample points per iteration and $K$ is the number of shots taken per sample point. Each data point represents the mean of 50 independent optimization runs on one representative problem instance, represented by an Erd\H{o}s-R\'enyi random unweighted graph. The initial parameter choice, $\btheta^{(0)}$ is the same for all runs; however, the sample points on each patch obtained via Latin hypercube sampling are chosen independently for each run. Error bars indicate standard error of the mean. Additional details on hyperparameter choices and implementation notes can be found in \cref{app:notes}.    }
\label{fig:qaoa}
\end{figure*}

\subsection{Quantum Approximate Optimization Algorithm}
\label{sec:qaoa}
The quantum approximate optimization algorithm (QAOA) is a variational circuit approach to combinatorial optimization \cite{Farhi_QAOA_2014}, where the optimization problem is encoded in a \emph{problem Hamiltonian}, $\h{H}_p$, whose ground state encodes the solution to the problem.
A commonly studied example is the MaxCut problem, which aims to partition an $n$-node graph into two sets of nodes, such that the weight of the edges going between the partitions is maximized. A MaxCut problem instance is encoded in an $n$-qubit Ising Hamiltonian of the form $\h{H}_p = \sum_{(i,j)\in \mathcal{E}} w_{ij} \h{Z}_i\h{Z}_j$, where $\h{Z}_i$ is a Pauli $Z$ matrix on qubit $i$ tensored with the identity on all other qubits, and $\mathcal{E}$ is the set of edges in the graph, each with weight $w_{ij} \in \mathbb{R}$.

QAOA approaches the goal of preparing low energy eigenstates of $\h{H}_p$ by $p$ iterated applications of a two-layer ansatz to a product input state to produce the output state:
\begin{align}
	\ket{\psi(\bgamma, \bbeta)} = \prod_{l=1}^p e^{-i\beta_p \h{H}_d}e^{-i\gamma_p \h{H}_p} \ket{+}^{\otimes n},
\end{align}
where $\ket{+} = \nicefrac{1}{\sqrt{2}}(\ket{0}+\ket{1})$, and $\h{H}_d = \sum_{i=1}^n \h{X}_i$. The variational parameters $\bgamma=(\gamma_1,..., \gamma_p)$ and $\bbeta = (\beta_1, ..., \beta_p)$ are optimized such that the energy of the output state is minimized; \ie the objective function is $V_{QAOA}(\bgamma, \bbeta) = \bra{\psi(\bgamma, \bbeta)}\h{H}_p \ket{\psi(\bgamma,\bbeta)}$. The \emph{approximation ratio}, which quantifies how close to the true ground state any $\ket{\psi(\bgamma, \bbeta)}$ is, is defined by $r = V_{QAOA}/E_0$, where $E_0$ is the true ground state energy of $\h{H}_p$.

If global optima to $V_{QAOA}$ can be found, in the $p\to\infty$ limit QAOA prepares the ground state of $\h{H}_p$, which encodes the solution to the original combinatorial optimization problem \cite{Farhi_QAOA_2014}. Moreover, in this case $r$ increases monotonically with $p$, although the question of what $p$ is required for $r$ to surpass approximation ratios achievable by classical approximation methods is an open one. It is clear that the variational optimization, and even finding good quality local minima of $V_{QAOA}$, becomes challenging with increasing $p$.

In terms of the parameters defined in the general description of VQAs in \cref{sec:approach}, it is important to note that the QAOA objective is defined through an observable, $H_p$, that only consists of commuting terms. Therefore, one only needs to measure in the computational basis for QAOA, meaning that we have $\nu=1$ and we require $\mathcal{K} = K$ total shots per sample point.


\cref{fig:qaoa} shows the results of simulated L-BFGS-B, SPSA, and SBO optimization runs of QAOA applied to MaxCut instances on (Erd\H{o}s-R\'enyi) random unweighted graphs of (a) $n=4$ vertices using $p=2$ layers, (b) $n=6$ vertices using $p=4$ layers, and (c) $n=10$ vertices using $p=7$ layers.
We plot the results of each run using $K \tau = 5000$ shots per iteration. We repeated these tests with various values of $K \tau$ ranging from $10^3$ to $10^5$ and observed qualitatively similar results.

We choose L-BFGS-B here as an example of a gradient-free optimizer. We use a gradient-free optimizer because the gradient of our noisy objective function is not directly available and therefore traditional gradient descent cannot be used. We found that L-BFGS-B, although it still performs very poorly due to the noisy objective function, significantly outperformed Nelder-Mead, another widely-used gradient-free optimizer.

At the smallest problem size, SBO achieves a lower error than SPSA for up to $M \sim 10^3$ iterations, indicating that it converges on a good approximation of the local minimum more efficiently. At the larger problem sizes, SBO achieves a lower error than SPSA for even larger numbers of iterations. For perspective, we note that an optimization run with $K \tau = 5000$ shots per iteration and $M = 10^3$ iterations would require a total of $K \tau M = 5 \times 10^6$ experimental shots. This would require an experimental duration on the order of several minutes using a typical superconducting QIP, or on the order of several days using a typical trapped-ion QIP. Because our results indicate that SBO significantly outperforms SPSA in this regime, it appears likely that SBO will achieve lower error than SPSA for many QAOA experiments that can be realistically implemented on current and near-future devices.

\subsection{Variational Quantum Eigensolver}
\label{sec:vqe}
The first VQA was the so-called variational quantum eigensolver (VQE) \cite{Peruzzo_2013}, which aims to prepare the ground state of an $n$-qubit Hamiltonian,  $\h{H}_E$, that encodes the energy of a molecule. The variational circuits and parameters, $\btheta$, that prepare candidate states vary according to the wavefunction \emph{ansatz} that is used \cite{McClean_2016}. In all cases, the objective function is defined as $V_{VQE}(\btheta) = \bra{\psi(\btheta)} \h{H}_E \ket{\psi(\btheta)}$. In general, $\nu>1$ for nontrivial $\h{H}_E$ and hence measurements in multiple bases are necessary.

\begin{figure*}[t]
\centering
  (a)
  \raisebox{-0.95\height}{
    \includegraphics[width=.8\columnwidth]{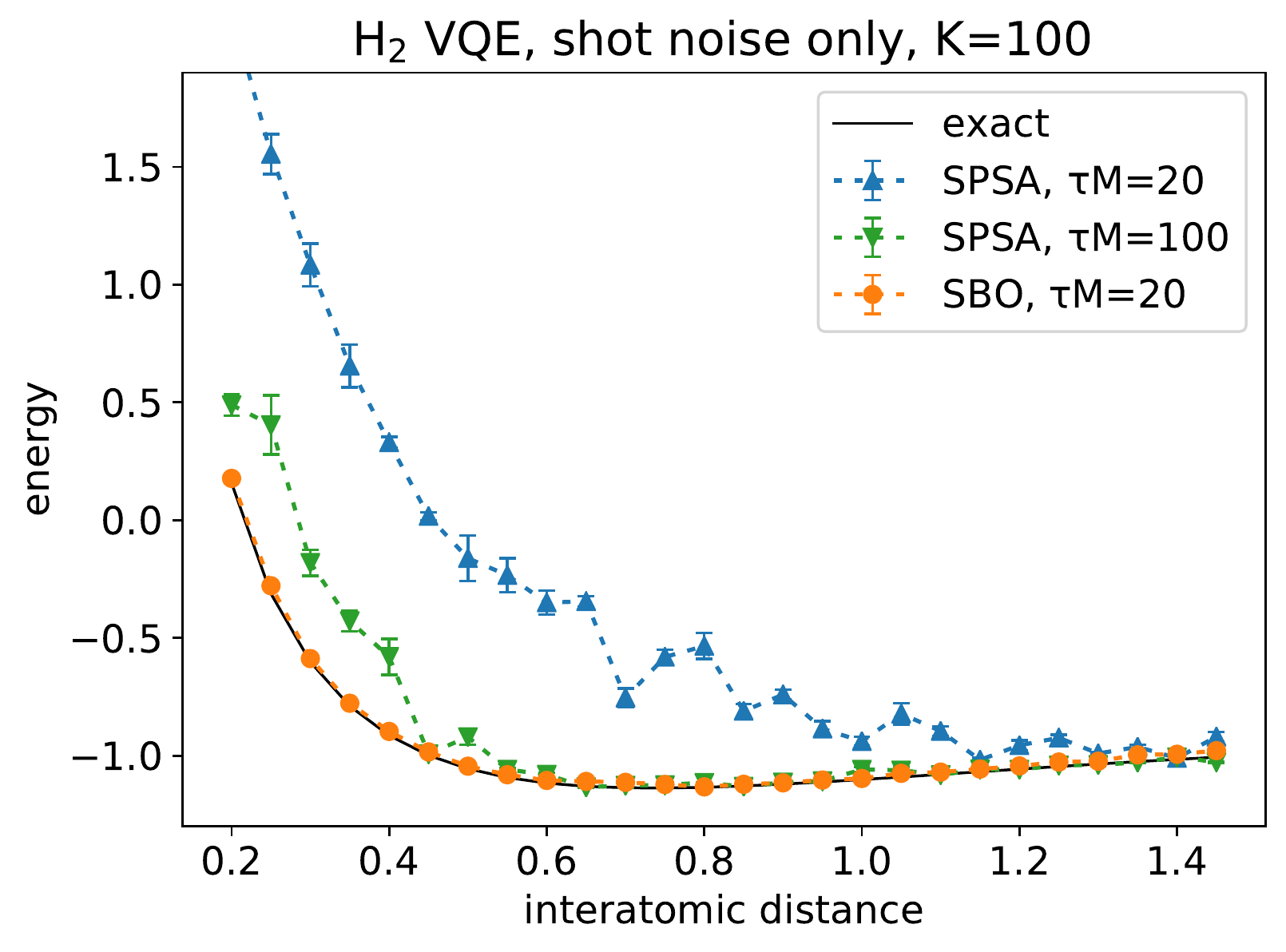}
  }
  \hspace{.1\columnwidth}
  (b)
  \raisebox{-0.95\height}{
    \includegraphics[width=.8\columnwidth]{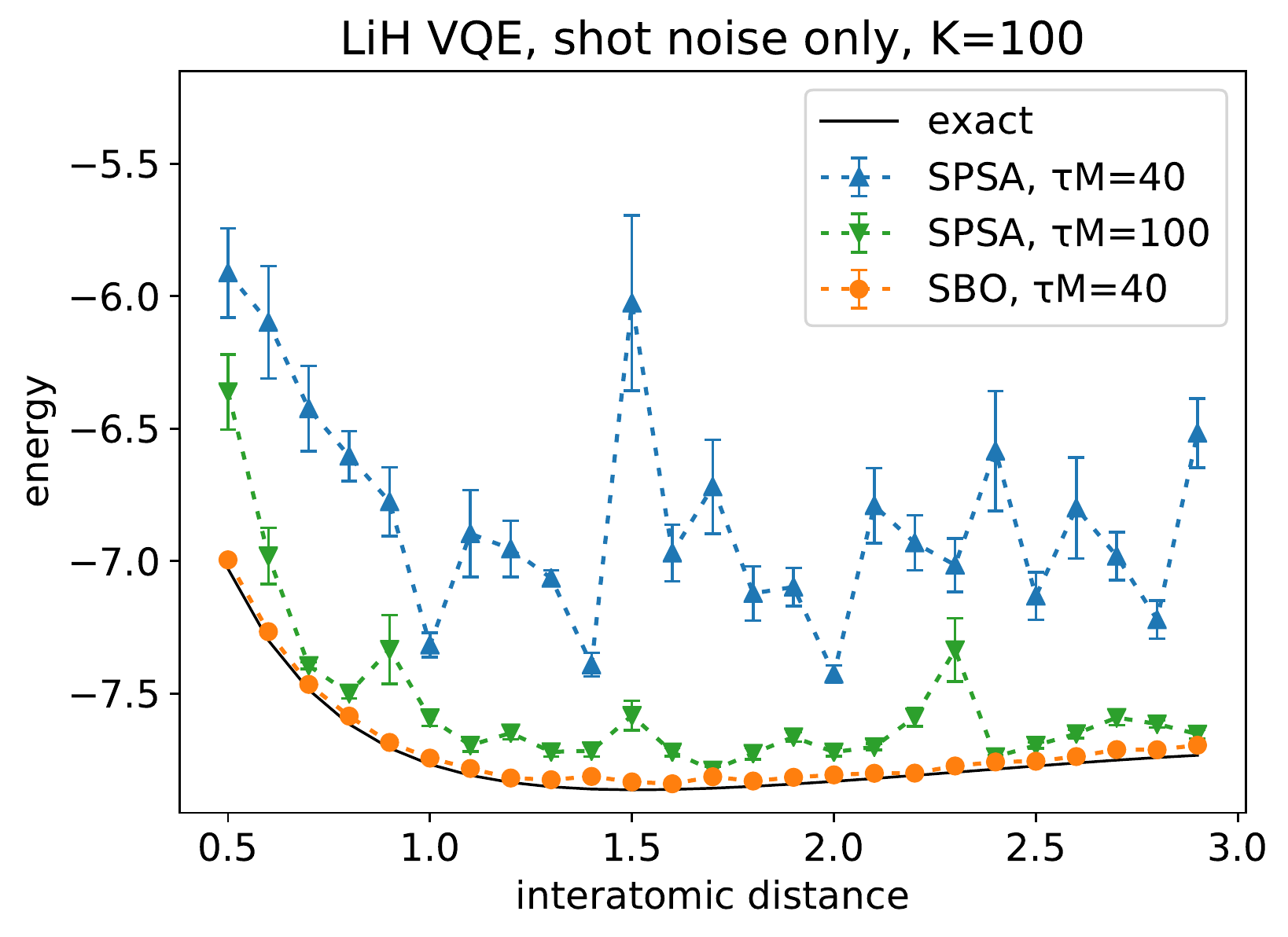}
  } \\
  (c)
  \raisebox{-0.95\height}{
    \includegraphics[width=.8\columnwidth]{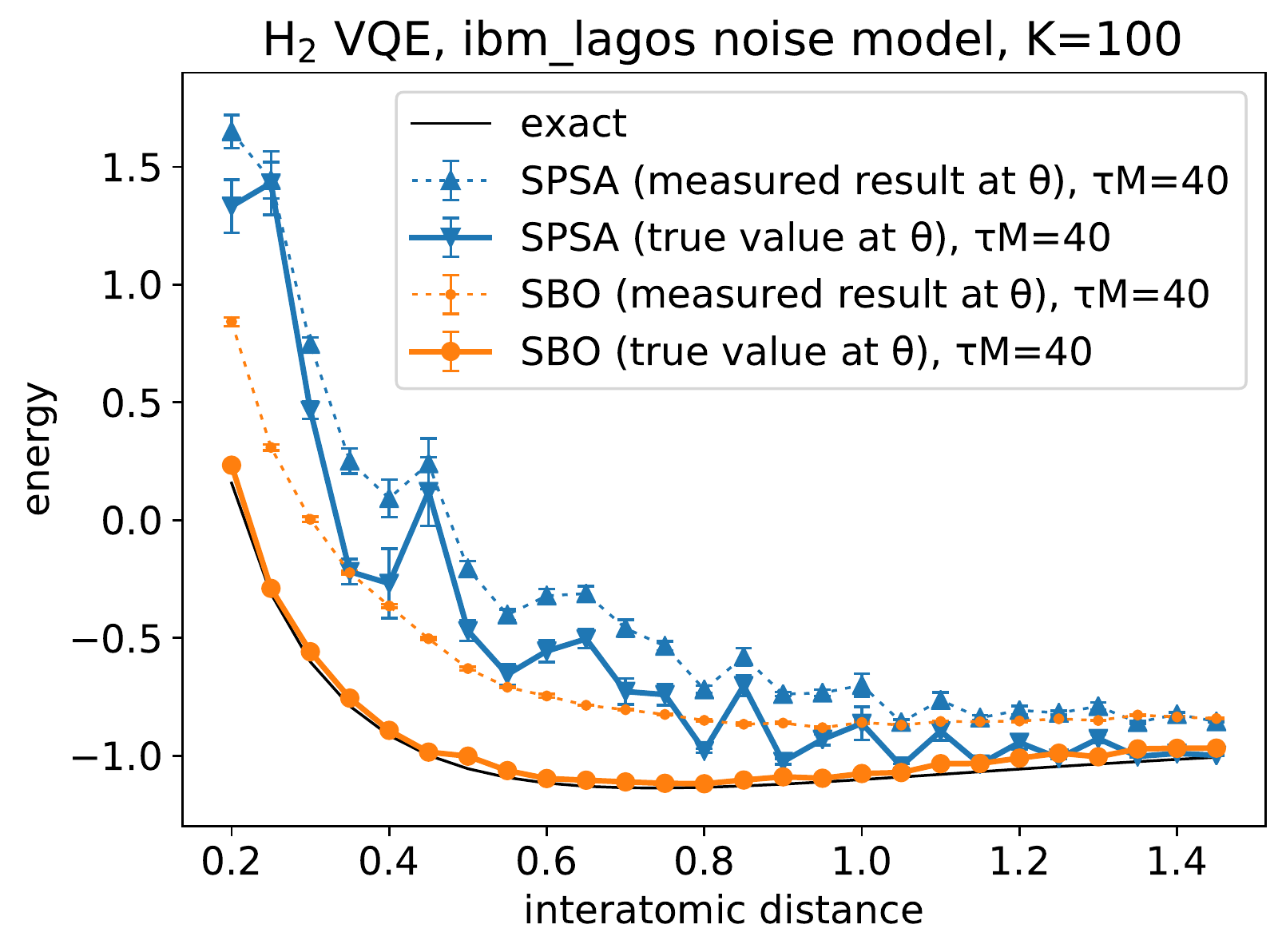}
  }
  \hspace{.1\columnwidth}
  (d)
  \raisebox{-0.95\height}{
    \includegraphics[width=.8\columnwidth]{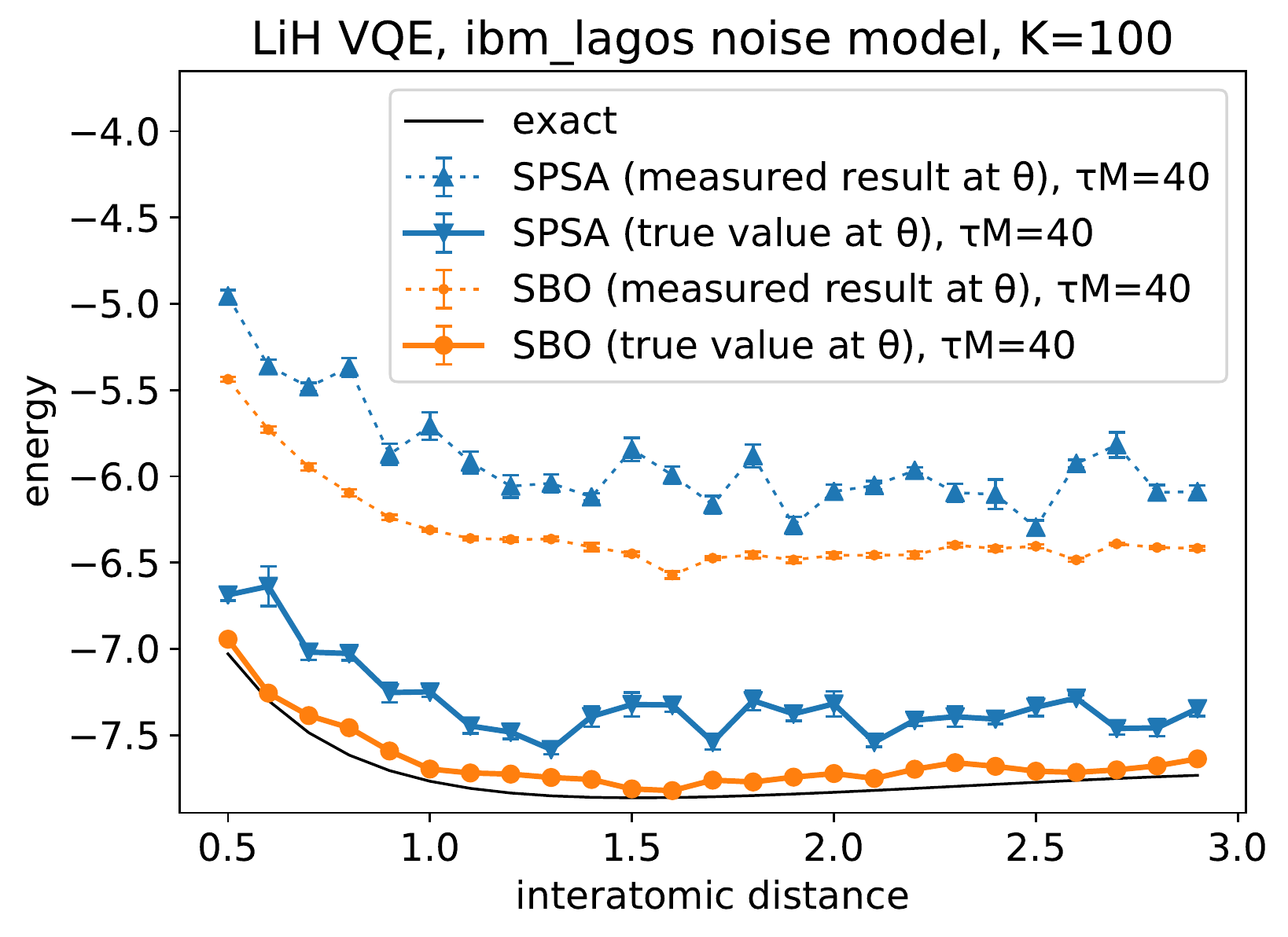}
  }
\caption{
    Performance of SPSA and Gaussian kernel-based SBO optimization runs on common small-scale VQE problems using simulators with and without realistic hardware noise.
    Each plot displays the minimum energy obtained for each bond length under the specified optimization conditions.
    The $x$-axis represents the interatomic distance (in angstroms) used for the energy calculation. The $y$-axis represents the energy value (in hartrees) obtained at the conclusion of each optimization run.
    $\tau M$ is the total number of energy measurements performed in the optimization run, where $\tau$ is the number of sample points per iteration and $M$ is the number of optimization iterations. To measure the energy, $K=100$ shots are taken per measurement basis per sample point. Each data point represents the mean of five independent optimization runs at the given setting. Error bars indicate standard error of the mean.
    On each plot, a solid black curve indicates the exact minimum energy value for the given setting.
    In (a) and (b), we use an ideal simulator which has only shot noise and no other errors. The data points connected by dashed curves represent the energy value obtained by evaluating the ansatz on the ideal simulator at the found optimal parameter values $\btheta$.
    In (c) and (d), we use a noisy simulator implementing a typical noise model and coupling map obtained from the seven-qubit IBM Q Lagos device. The data points connected by dashed curves represent the energy value obtained by evaluating the ansatz on the noisy simulator at the found optimal parameter values $\btheta$. The data points connected by solid curves represent the energy value obtained by evaluating the ansatz at $\btheta$ using an ideal simulator. Additional details on hyperparameter choices and implementation notes can be found in \cref{app:notes}.}
\label{fig:vqe}
\end{figure*}

\cref{fig:vqe} shows the results of simulated SPSA and SBO optimization runs of VQE for estimating the ground state energy of H$_2$ and LiH molecules at various interatomic bond lengths using the unitary coupled cluster ansatz with Hartree-Fock initial state. The H$_2$ ansatz uses four qubits and $|\btheta| = 3$ variational parameters, while the LiH ansatz uses four qubits and $|\btheta| = 8$ variational parameters.
In \cref{fig:vqe} (a) and (b), under shot noise only, we observe that SBO produces a much more accurate estimate of the ground state energy than SPSA using the same number of energy measurements $\tau M$, and it remains more accurate even than using SPSA with $\tau M$ increased by a factor of 2.5 to 5.
In \cref{fig:vqe} (c) and (d), using a simulator with a realistic hardware noise model, we observe that SBO achieves consistently lower estimates of the ground state energy than SPSA. In addition, by taking the parameters $\btheta$ found by the noisy optimization runs and evaluating the ansatz with those values on an ideal simulator, we observe that the parameter values obtained by SBO correspond to energy values which are much closer to the exact ground state than those obtained by SPSA.

\section{Discussion}
\label{sec:disc}

From both the QAOA and VQE illustrations in \cref{sec:eg}, we observe that SBO often achieves a lower error than SPSA for an equivalent number of iterations or experimental shots. From the QAOA results in \cref{fig:qaoa}, we note that this advantage tends to become more pronounced as the problem complexity increases. We believe these results are a good indication that, for many near-term applications, SBO will achieve better variational parameter estimates with fewer experimental repetitions than existing techniques such as SPSA. Additionally, because the surrogate function smooths out shot noise, SBO often requires fewer shots per sample point than SPSA to produce a result that is equivalent or better.

One unique feature of SBO is that each iteration requires taking samples for many different parameter settings, as opposed to a technique like SPSA which uses only two sample points per iteration. This may provide a particular advantage for experimental platforms that suffer a high latency cost from loading new circuits between each optimization iteration. If the system can program and execute an entire batch of circuits without paying this latency cost between each circuit, this may provide an additional speed advantage, as well as increased robustness against drift in experimental parameters.

Finding global optima of parametrized quantum circuits often suffers from the problem of
``barren plateaus'' \cite{McClean_2016}, wherein the objective function, $V(\btheta)$, exhibits exponentially-vanishing gradients, both in the absence and presence of hardware noise, making optimization exceeding challenging. Some techniques around this problem are to formulate \emph{local cost functions} \cite{cerezo2021cost} and to utilize variational circuit forms that do not exhibit barren plateaus \cite{Grimsley_2022}. We emphasize that SBO is not a technique to address the problem of barren plateaus. Instead, it is an approach to increase the performance of classical optimization loops and to reduce the experimental burden in the VQA setting. These issues are orthogonal to the barren plateaus issue -- strategies to construct variational circuits that do not possess barren plateaus \emph{and} the use of more advanced classical optimization techniques like SBO will be critical for scaling VQAs.

A promising avenue for future work is the application of more powerful surrogate models to the VQA setting, \eg neural network-based methods for approximation \cite{Lee_2021} might have better rates of convergence with limited experimental data. In addition, building surrogate models to not only perform smoothing and approximation, as we have done here, but also physics-informed error mitigation to counter decoherence is a potentially fruitful direction. 

\pagebreak
\section*{Code Availability}

A freely-available Python implementation of the Gaussian kernel-based SBO optimizer, including examples of integration with IBM's Qiskit library, is available at \url{https://github.com/sandialabs/sbovqaopt}.

\section*{Acknowledgements}
This work was supported by the U.S. Department of Energy, Office of Science, Office of Advanced Scientific Computing Research, under the Quantum Computing Application Teams program.
R.S. was also supported by NSF award DMR-1747426.
M.S. was also supported by the U.S. Department of Energy, Office of Science, National Quantum Information Science Research Centers.
Sandia National Laboratories is a multimission laboratory managed and operated by NTESS, LLC., a wholly owned subsidiary of Honeywell International, Inc., for the U.S. DOE's NNSA under contract DE-NA-0003525. This paper describes objective technical results and analysis. Any subjective views or opinions that might be expressed in the paper do not necessarily represent the views of the U.S. Department of Energy or the United States Government.

\bibliography{surrogate}

\clearpage
\widetext
\appendix

\section{Bound on number of critical points in a variational cost function}
\label{app:bound}
As discussed in the main text, a conservative heuristic for choosing the SBO patch size, $\ell$, is to choose it such that there are $O(1)$ critical points in the variational cost function within a $\ell^D$ hypercube. In the following, we will formulate a bound on the total number of critical points in general variational cost functions, $N_{\rm crit}$, in terms of the key parameters in a variational circuit: $n$, the number of qubits, and $D$, the number of variational parameters. If we then assume that these critical points are distributed uniformly in parameter space, we require
\begin{align}
	\left(\frac{\ell}{2\pi}\right)^D N_{\rm crit} \sim 1 \quad \Rightarrow \quad \ell \sim N_{\rm crit}^{-\nicefrac{1}{D}}.
	\label{eq:scaling}
\end{align}

First, we write a general variational cost function as
\begin{align}
	V(\btheta) = \tr\left[\h{O}\left(\prod_{j=1}^D e^{-i\theta_j \h{H}_j}\right) \rho_0 \left(\prod_{j=1}^De^{i\theta_j \h{H}_j}\right) \right],
	\label{eq:gen_V}
\end{align}
where $\h{H}_j$ are the $n$-qubit Hamiltonians representing the variational ansatz. The $H_j$ are multi-qubit Hamiltonians in general, and to derive an informative bound on $N_{\rm crit}$ we should take into account the complexity in decomposing $e^{-i\theta\h{H}_j}$ into implementable unitaries. To make things concrete, we will work with a decomposition of the form:
\begin{align}
	e^{-i\theta_j\h{H}_j} = \h{U}^{(j)}_{\lambda_j+1} \h{Z}_{\iota(\lambda_j)}(\theta_j) \h{U}^{(j)}_{\lambda_j-1} ... \h{Z}_{\iota(1)}(\theta_j)\h{U}^{(j)}_{1},
\end{align}
where the $\h{U}^{(j)}_{i}$ are $\theta_j$-independent $n$-qubit unitaries, and $\h{Z}_{\iota(i)}(\theta_j)$ is a $Z$ rotation on qubit $\iota(i)$. This decomposition implements $e^{-i\theta \h{H}_j}$ with $\lambda_j$ rotations by the variational parameter $\theta_j$, and we think of $\lambda_j$ as parametrizing the complexity of $\h{H}_j$. Note that $\lambda_j$ can have a dependence on $n$ since it is the complexity of decomposing an $n$-qubit unitary. For practical quantum computations, $\lambda_j = \mathcal{O}(\text{poly}(n))$. The decomposition above is not unique, but we note that it is experimentally relevant since in many modern quantum computing architectures the only variable angle gates are single qubit $\h{Z}(\theta)$ rotations.

As an example, consider the decompositions of the two terms in a layer of the QAOA ansatz:
\begin{align}
	e^{-i\beta \sum_{i=1}^n \h{X}_i} &= \h{H}^{\otimes n} \h{Z}(\beta)^{\otimes n} \h{H}^{\otimes n},  \\
	e^{-i\gamma \sum_{(i,j)\in \mathcal{E}} w_{ij}\h{Z}_i\h{Z}_j} &= \prod_{k=1}^{|\mathcal{E}|} \textsc{cnot}_{i_k, j_k} \h{Z}_{j_k}(w_{i_kj_k}\gamma) \textsc{cnot}_{i_k, j_k},
\end{align}
where $\h{H}$ in the first line is a Hadamard gate, and in the second line $i_k$ and $j_k$ index the nodes that edge $k$ connects. $\lambda=1$ in the first line, and in the second line $\lambda\leq|\mathcal{E}|$. The exact $\lambda$ for a decomposition of the $\h{Z}\h{Z}$ interactions will depend on the QAOA problem graph and which \textsc{cnot} gates can be executed in parallel -- since a \textsc{cnot}$_{ij}$ and a \textsc{cnot}$_{jm}$ cannot be simultaneously applied, a node $j$ that has edges to both node $i$ and node $m$ will need its $\h{Z}\h{Z}$ interactions implemented sequentially (even assuming full connectivity in the hardware). In general, for a $\kappa$-regular problem graph, $\lambda = \kappa$ if the device is fully connected (\ie a $\h{Z}\h{Z}$ gate can be implemented between all qubits connected by an edge in $\mathcal{E}$).   

Returning to the general variational cost function in \cref{eq:gen_V} and substituting the compiled form of each unitary, we get
\begin{align}
	V(\btheta) = \tr\left[\h{O} \left(\prod_{j=1}^D \h{U}^{(j)}_{\lambda_j+1}\prod_{k=1}^{\lambda_j} \h{Z}_{\iota(k)}(\theta_j) \h{U}^{(j)}_k \right) \rho_0 \left(\prod_{j=1}^D\prod_{k=1}^{\lambda_j} \h{U}^{(j)\dagger}_k \h{Z}_{\iota(k)}(-\theta_j) \h{U}^{(j)\dagger}_{\lambda_j+1}\right)\right].
\end{align}
Finally, without loss of generality taking $\rho_0=\ket{0}\bra{0}$ (where $\ket{0}$ is shorthand for the $n$-qubit state $\ket{0}^{\otimes n}$), we write
\begin{align}
	V(\btheta) = \bra{0} \prod_{j=1}^D\prod_{k=1}^{\lambda_j} \h{U}^{(j)\dagger}_k \h{Z}_{\iota(k)}(-\theta_j) \h{U}^{(j)\dagger}_{\lambda_j+1} ~\h{O}~ \prod_{j=1}^D \h{U}^{(j)}_{\lambda_j+1}\prod_{k=1}^{\lambda_j} \h{Z}_{\iota(k)}(\theta_j) \h{U}^{(j)}_k \ket{0} = \bra{0}\prod_{j=1}^D\mathcal{U}_j\dg~\h{O}~\prod_{j=1}^D\mathcal{U}_j\ket{0},
\end{align}
where $\mathcal{U}_j \equiv \h{U}^{(j)}_{\lambda_j+1}\prod_{k=1}^{\lambda_j} \h{Z}_{\iota(k)}(\theta_j) \h{U}^{(j)}_k$ are the decompositions. Taking the derivative with respect to one of the angles, we get
\begin{align}
	\frac{\partial V(\btheta)}{\partial \theta_t} &= \sum_{r=1}^{\lambda_t} \bra{0} \left[\prod_{j=1}^{t-1} \mathcal{U}\dg_j\right]  \left(\prod_{k=1}^{r-1} \h{U}^{(t)\dagger}_k \h{Z}_{\iota(k)}(-\theta_t)\right)\h{U}^{(t)\dagger}_r \h{Z}_{\iota(r)}(-(\nicefrac{\pi}{2}+\theta_t)) \left(\prod_{k=r+1}^{\lambda_t} \h{U}^{(t)\dagger}_k \h{Z}_{\iota(k)}(-\theta_t)\right)\h{U}^{(t)\dagger}_{\lambda_t+1} \left[\prod_{j=t+1}^{D} \mathcal{U}\dg_j\right] \nn \\
	&\quad\quad\quad\quad~\h{O}~ \left[\prod_{j=1}^D \mathcal{U}_j\right] \ket{0} ~+~ c.c,
	\label{eq:dV}
\end{align}
since $\nicefrac{\partial}{\partial \theta}\,\h{Z}(-\theta)=e^{i(\nicefrac{\pi}{2}+\theta) \h{Z}}$.  

Since all the dependence of this expression on the angles $\btheta$ are within the $\h{Z}$ rotations, and $\h{Z}(\theta)=\cos(\theta)\h{I} + i \sin(\theta)\h{Z}$, we conclude that $\nicefrac{\partial V(\btheta)}{\partial \theta_t}$, and $V(\btheta)$ for that matter, are trigonometric polynomials in the angles. Since taking the derivative of $V(\btheta)$ does not introduce any new $\h{Z}$ rotations, and only shifts the angle of some of the rotations in $V(\btheta)$, the maximum degree of this trigonometric polynomial is the same for $\nicefrac{\partial V(\btheta)}{\partial \theta_t}$ and $V(\btheta)$. 

Now we write each decomposition more explicitly as a trigonometric polynomial, using the fact that $\h{U} \h{Z}(\theta) \h{U}\dg = \cos(\theta)\h{I} -i\sin(\theta)\h{U}\h{Z}\h{U}\dg$:
\begin{align}
	\mathcal{U}_j = \h{\Lambda}^{(j)}_{\lambda_j+1}\prod_{k=1}^{\lambda_j}\left(\cos(\theta_j)\h{I} - i\sin(\theta_j)\h{\Lambda}^{(j)}_{k}\right) = \sum_{\{\alpha,\beta\geq 1 :\alpha+\beta=\lambda_j\}} \cos^{\alpha}(\theta_j)\sin^{\beta}(\theta_j) \h{\Gamma}^{(j)}_{\alpha,\beta},
\end{align}
where $\h{\Lambda}_k^{(j)}= \h{U}^{(j)\dagger}_1...\h{U}^{(j)\dagger}_k \h{Z}_{\iota(k)} \h{U}^{(j)}_k ... \h{U}^{(j)}_1$ for $k\leq \lambda_j$,  $\h{\Lambda}_{\lambda_j+1}^{(j)}= \prod_{s=1}^{\lambda_j+1} \h{U}^{(j)}_s$, and $\h{\Gamma}^{(j)}_{\alpha,\beta}$ is an operator that is a multiple of some of the $\h{\Lambda}_k^{(j)}$ that we do not need to specify. Thus $\mathcal{U}_j$ is a trigonometric polynomial with maximum degree $\lambda_j$ and operator coefficients. Using the same argument for all the $\mathcal{U}_j$, we can write $V(\btheta)$ as:
\begin{align}
	V(\btheta) &= \bra{0}\prod_{j=1}^D \left(\sum_{\{\alpha_j,\beta_j\geq1:\alpha_j+\beta_j=\lambda_j\}}\cos^{\alpha_j}(\theta_j)\sin^{\beta_j}(\theta_j) \h{\Gamma}^{(j)\dagger}_{\alpha_j,\beta_j} \right) \h{O} \prod_{j=1}^D \left(\sum_{\{\alpha_j,\beta_j\geq1:\alpha_j+\beta_j=\lambda_j\}}\cos^{\alpha_j}(\theta_j)\sin^{\beta_j}(\theta_j) \h{\Gamma}^{(j)}_{\alpha_j,\beta_j} \right) \ket{0} \nn \\
	&= \sum_{\{\alpha_j,\beta_j\geq1:\alpha_j+\beta_j=2\lambda_j\}}\prod_{j=1}^D \cos^{\alpha_j}(\theta_j)\sin^{\beta_j}(\theta_j) g^{(j)}_{\alpha_j,\beta_j},
	\label{eq:D_eqns}
\end{align}
where in the final line $g^{(j)}_{\alpha_j,\beta_j} \in \mathbb{R}$. Not all of the terms in this sum will be present since $g^{(j)}_{\alpha_j,\beta_j}$ could be zero. However, without further assumptions about the problem, we must assume they are all present. In that case, this is a trigonometric polynomial with maximum degree $d=\sum_{j=1}^D 2\lambda_j$. And as argued above, all derivatives of $V(\btheta)$ are trigonometric polynomials with the same maximum degree. Therefore, critical points of the variational cost function are defined by a set of $D$ trigonometric polynomial equations in $D$ angle variables; \ie $\nicefrac{\partial V(\btheta)}{\partial \theta_t}=0$ for all $t$. 

To count the number of critical points, we wish to count the number of solutions to this system of equations. We are unaware of any applicable bounds on the number of solutions of such trigonometric polynomial systems, and therefore proceed by transforming this into a system of standard polynomials. To do so, we introduce new variables, $s_j = \sin(\theta_j)$, and $c_j=\cos(\theta_j)$, $1\leq j \leq D$, which transforms \cref{eq:D_eqns} into a degree $d$ polynomial equation in $2D$ variables. Hence in these two variables, the system $\nicefrac{\partial V(\btheta)}{\partial \theta_t}=0$ for all $t$, is a system of $D$ polynomial equations in $2D$ variables, with each equation being degree $d$. We supplement these with the equations encoding the constraint between $c_j$ and $s_j$, namely $c_j^2+s_j^2-1=0$, for $1\leq j \leq D$, to arrive at a system of $2D$ polynomial equations in $2D$ variables. $D$ of these equations have degree $d$, and $D$ of them (the constraints equations) have degree 2.

Now that we have a system of polynomial equations we can formulate a bound on the number of solutions to this system. For this purpose, we use B\'ezout's theorem, which states that in general, the number of common zeros for a set of $n$ polynomials in $n$ variables in $\mathbb{C}^n$ is given by the product of the degrees of the polynomials \cite{Penchevre_2016}. Applying this, we arrive at our bound for the number of critical points in $V(\btheta)$:
\begin{align}
	N_{\rm crit} \leq (2d)^D = \left(4\sum_{j=1}^D\lambda_j\right)^D
	\label{eq:bound}
\end{align}
In cases where $\lambda_j=\lambda$ for all $j$, $N_{\rm crit}\leq (4\lambda D)^D$.

We pause to emphasize that this is a particularly loose bound. Firstly, it is well-known that the bound provided by B\'ezout's theorem is loose. Compounding this, B\'ezout's theorem counts the number of zeros over $\mathbb{C}^{2D}$, whereas we are concerned with zeros over the domain $[-1,1]^{2D}$. Thus, although we do not expect this bound to be tight, it does highlight some useful parameter dependencies: (i) there is an exponential dependence on the number of parameters $D$, and (ii) the only dependence on $n$ is through the ansatz complexity $\lambda_j$.

Returning to the scaling of the patch size parameters, $\ell$, \cref{eq:scaling}, we arrive at:
\begin{align}
	\ell \gtrsim \left(4\sum_{j=1}^D \lambda_j\right)^{-1},
\end{align}
which, taking into account $\lambda_j=\mathcal{O}(\text{poly}(n))$, results in the scaling $\ell = \Omega(\nicefrac{1}{\text{poly}(D,n)})$.

As an example, consider QAOA on $\kappa$-regular graphs. For a QAOA variational ansatz with $p$ layers, $D=2p$, and as discussed above, $\lambda_j$ alternates between $1$ and $\kappa$. Therefore, for this example we get $\ell \gtrsim \left(4p(\kappa +1)\right)^{-1}$. Note that since the $\lambda_j$ have no dependence on $n$ in this case, we get $n$-independent scaling.

\begin{figure*}[t!]
\centering
  \includegraphics[width=1.0\columnwidth]{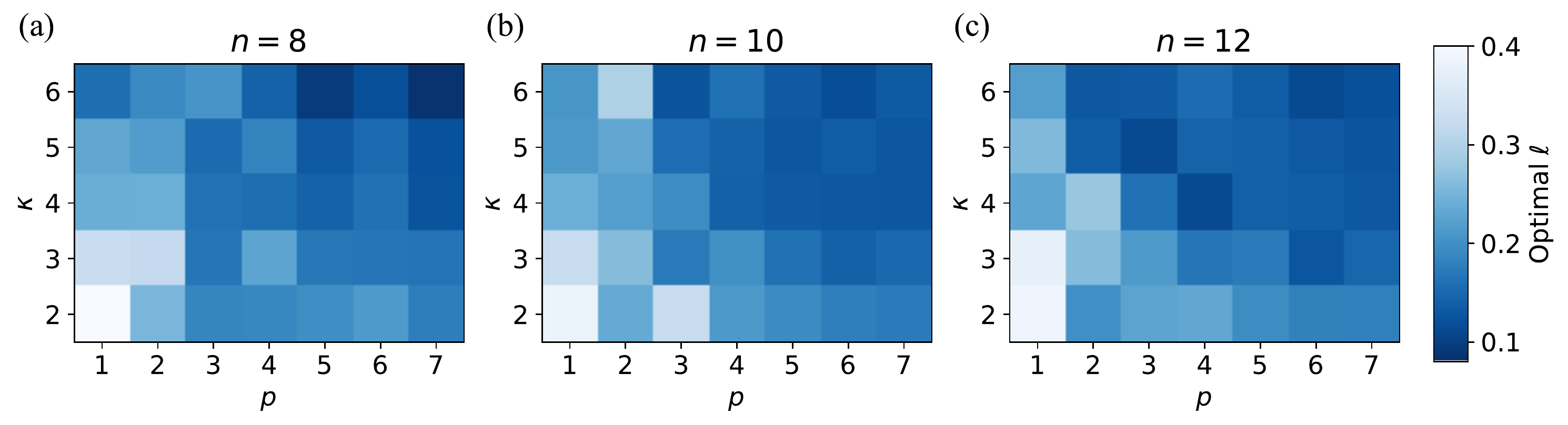}
\caption{
    Numerical estimation of the optimal SBO patch size $\ell$ for various instances of $n$-qubit, $p$-layer MaxCut QAOA on randomly-generated $\kappa$-regular graphs.
    Each data point is obtained by averaging the final error of 10 independent SBO runs for each $\ell \in \{0.02, 0.04, \dots, 0.40\}$, and then using cubic splines to fit the results and find the value of $\ell$ which minimizes the average error. Each run uses $\tau=30$ sample points per patch, $K=60$ measurement shots per sample point, and $M=100$ optimization iterations.}
\label{fig:optimal-patch-size-scaling}
\end{figure*}

\begin{figure*}[t!]
\centering
  \includegraphics[width=0.8\columnwidth]{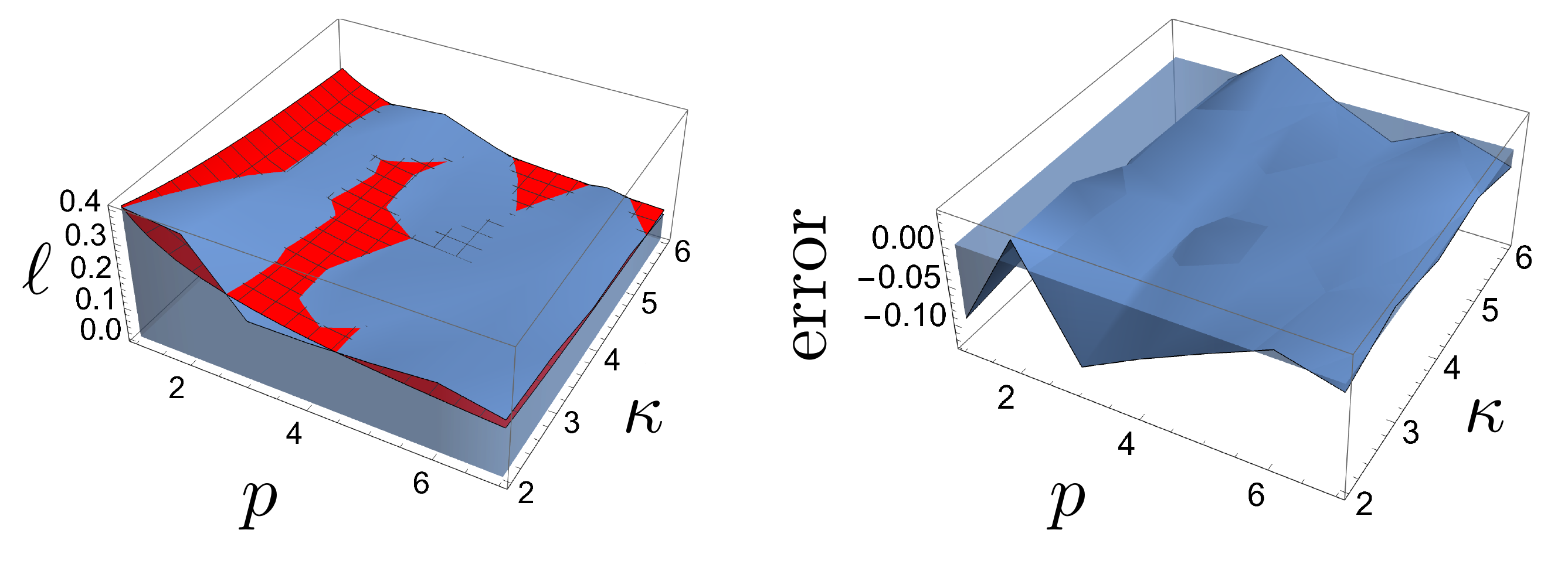}
\caption{
    Fit (left) and error in fit (right) of data in \cref{fig:optimal-patch-size-scaling}(a) to the functional form shown in \cref{eq:fit_form}, with fitting parameters $\alpha=0.5, \beta=0.7$.}
\label{fig:fit}
\end{figure*}

In \cref{fig:optimal-patch-size-scaling} we present numerically-determined optimal $\ell$ for QAOA on $\kappa$-regular graphs as we vary the relevant parameters: the number of QAOA layers $p$ (where $D=2p$), the number of qubits $n$, and the graph regularity $\kappa$. The independence of $\ell$ from $n$ is supported by this data, as the variation of the surfaces is negligible as $n$ is varied. In order to test the scaling prediction above, we fit the data in \cref{fig:optimal-patch-size-scaling}(a) to a functional form
\begin{align}
	\ell = \beta (\kappa p+\kappa)^{-\alpha},
	\label{eq:fit_form}
\end{align}
where the parameters $\alpha,\beta$ allow for numerical factors in the relations \cref{eq:scaling} and \cref{eq:bound}, and accommodate for the looseness of the bound in \cref{eq:bound}. The fit of this form to the $n=8$ data, along with the error of the fit, is shown in \cref{fig:fit}.

\section{Implementation notes for simulations in Sec. \ref{sec:eg}}
\label{app:notes}
Here we collect the implementation notes and hyperparameter choices for the simulations presented in the main text.

\subsection{Quantum Approximate Optimization Algorithm}
The QAOA circuit simulations were implemented using the pyQAOA package from \url{https://github.com/gregvw/pyQAOA}.
SPSA was implemented using the noisyopt package from \url{https://github.com/andim/noisyopt}.
L-BFGS-B was implemented using the \texttt{optimize.minimize} function in scipy.

\textbf{Simulation hyperparameters:} For \cref{fig:qaoa}, we chose hyperparameters by manual scans to optimize the performance of both SBO and SPSA on these problems. We use $\tau = 20$ for SBO, while $\tau = 2$ for SPSA by definition.
    In \cref{fig:qaoa}(a,b,c), we use SBO parameter $\ell = (0.2, 0.2, 0.1)$ and SPSA parameter $a = (0.2, 0.2, 0.1)$, respectively. We use SPSA parameters $c=0.2$, $\alpha=0.602$, and $\gamma=0.101$ for all simulations presented in this figure.

\subsection{Variational Quantum Eigensolver}
VQE simulations were implemented via qiskit using unitary coupled cluster (UCC) ansatz with Hartree-Fock initial state, following the procedure described at \url{https://qiskit.org/documentation/nature/tutorials/03_ground_state_solvers.html}.
For SPSA, we used the implementation provided by qiskit at \url{https://qiskit.org/documentation/stubs/qiskit.algorithms.optimizers.SPSA.html}, which includes automatic hyperparameter calibration.

\textbf{Simulation hyperparameters:} In \cref{fig:vqe} (a,b,c,d), we used $\tau = (4,5,8,10)$ and $\ell = (0.15, 0.1, 0.15, 0.1)$ for SBO, respectively, while $\tau = 2$ for SPSA by definition.

\end{document}